\documentclass[twocolumn,showpacs,preprintnumbers,amsmath,amssymb]{revtex4}
\usepackage{graphicx}
\usepackage{dcolumn}
\usepackage{bm}
\begin{document}
\title{Valence Bond Order and Antiferromagnetism in Silicene \\- ab initio Results}
\author{R. Vidya$^{1}$}
\email{vidyar@annauniv.edu}
\author{G. Baskaran$^{2,3}$}
\address{$^{1}$Department of Medical Physics, Anna University, Guindy Campus, Chennai 600 025, India}
\address{$^{2}$Institute of Mathematical Sciences, CIT Campus, Taramani, Chennai 600 113, India}
\address{$^{3}$Perimeter Institute for Theoretical Physics, Waterloo, Ontario, Canada N2L 2Y5}
\date {\today}

\begin{abstract}
Silicene and Graphene are similar and have $\pi$-$\pi^*$ bands. However band width in silicene is only a third of graphene.
It results in a substantial increase in the ratio of Hubbard U to band width W, U/W $\sim$ 0.5 in graphene to $\sim 1$ in silicene. This enhancement, 2 dimensionality and phenomenology suggest a Mott insulator based ground state for silicene (G. Baskaran, arXiv:1309.2242). We lend support to the above proposal by showing, in an ab-initio calculation, that unlike graphene, silicene has two instabilities: i) a valence bond (Kekule) dimerization and ii) a weak two sublattice antiferromagnetic order. Presence of these instabilities, in the absence of fermi surface nesting, point to Mott localization, \textit{within the frame work of ab-initio scheme}. Substrate dependent structural reconstructions seen experimentally in silicene are interpreted as generalized Kekule bond order.
\end{abstract}

\maketitle

Extraordinary properties exhibited by graphene \cite{NovosolevGeim1,NovosolevGeim2,NovosolevGeim3}, has excited the community and encouraged them look for graphene analogues. A lot of attention has been paid on silicon based silicene \cite{slcnPrediction1,slcnPrediction2,slcnPrediction3,slcnReview}. Silicene is in some sense the  closest to graphene, as Si is just below carbon in the periodic table. Further, there is a possibility that silicene could be integrated into the existing silicon-based electronics industry. An intrinsic finite band gap in silicene, or a band gap that is easily opened by external perturbations, could offer promising advantage over graphene for broader applications, such as bipolar devices and high-performance field-effect transistors.

In this context, a proposal by one of us \cite{GBsilicene}, that the ground state of silicene is a small gap Mott insulator makes the silicene scenario attractive and exciting, from several points of view, including possibility of high Tc superconductivity on doping. Existing experimental results on silicene has been interpreted\cite{GBsilicene} in terms of Mott localization. Indeed, in an important recent experimental development \cite{Akinwande1} on silicene transistor fabrication, certain unusual phenomena also point to possible Mott physics in silicene.

Aim of the present article is to provide additional support for our Mott insulator scenario, by focussing on unusual and ubiquitous structural reconstructions seen in experiments, as compared to graphene. We present our ab-initio calculations and infer that a weaker p-$\pi$ bond in silicene, compared to graphene, favors i) Mott localization and ii) structural reconstructions in silicene. Lattice of silicene and substrate on which it is grown, take advantage of this and gain electron-lattice interaction energy through valence bond trapping, and an accompanying spatial patterns of bond length changes and puckering.

The article is organized as follows. We review quantum chemistry of silicene and some key experimental results that distinguish silicene from graphene. Then we summarize results of our state-of-the art ab-initio calculations: two dimensional valence bond (Kekule) order coexisting with two sublattice antiferromagnetic order with a small moment. Then we discuss possibility of substrate induced ordered topological defects such as zero charge spin-half soliton and charge $\pm$ e, spin-0 solitons in silicene. Some known reconstructions in silicene are discussed in terms of ordered lattice or solitons, valence bonds and resonating valence bond states.

Experimental synthesis of silicene is hard. Unlike graphene, it has not been possible to synthesize free standing silicene on insulating substrates. Silicene needs a \textit{metallic substrate as a stabilizing agent}. Silicene has been grown on on metallic substrates Ag, ZrB$_2$ and Ir \cite{slcnAg1,Vogt,slcnAg2,slcnZrB2,slcnIr,ChenSTS,slcnLandauLevel,NaDoping} with interesting ARPES and STM results. However Ag, a popular substrate modifies properties of silicene and leaves its imprints \cite{VogtDebate1,VogtDebate2,VogtDebate3,meng13,KuheinBansilARPES}. In contrast to the stiff and flat graphene, silicene exhibits a variety of structural reconstructions containing periodic patterns of bond length changes and $c$-axis puckering.

Eventhough silicon lies just below carbon in the periodic table its atomic radius is $\sim$  2.4 $\AA$, large compared to that of carbon $\sim$ 1.7 Au. This increase results in a significantly stretched Si-Si bond length $\sim$ 2.4 $\AA$, compared to $\sim$ 1.45 $\AA$ in graphene. This is primarily responsible for a nearly 3-fold reduction \cite{slcnPrediction1,slcnPrediction2,slcnPrediction3,slcnReview} in the hopping matrix element and width of $\pi$-$\pi^*$  bands. Added to the above is a small sp$^3$ mixing in silicene, arising from a smaller 3s-3p slitting $\sim$ 5 eV in Si, compared to a larger 2s-2p splitting $\sim$ 9 eV in carbon. In spite of the sp$^2$ mixing, according to electronic structure calculations, silicene band structure remains qualitatively the same as graphene, \textit{except for a reduction of $\pi$-$\pi^*$ band width by a factor of three}.

According to our theory this 3 fold band width reduction causes Mott localization. As we discussed in reference \cite{GBsilicene} in some detail, a body of existing experimental results are consistent with a Mott insulating ground state for silicene. One such phenomenology we focus in the present work is the ubiquity of structural reconstructions in all known silicene layers, grown on substrates. This is to be contrasted with graphene grown on substrates, where structural reconstructions hardly occur. A considerably weakened 3p-$\pi$ bonding reduces the Si-Si force/spring constants. A weaker bond (reduced hopping matrix element) in turn favors Mott localization as well as structural distortions.

An enhanced spin orbit coupling in silicene, compared to graphene, has attracted much attention in the community: spin-orbit coupling $\lambda_{\rm SO} \sim 3~ meV$ in silicene and $\lambda_{\rm SO} \sim \mu eV $ in graphene. However the spin-orbit coupling energy scale is much smaller compared to the scale of Mott-Hubbard gap and exchange couplings that we suggest for silicene; in the present work we will not consider spin-orbit coupling.

In what follows we report accurate ab-initio DFT-based calculations performed on structurally relaxed silicene, in non-magnetic (NM) and antiferromagnetic (AFM) spin states. We present computational details and details of electronic structure results in the supplement.

We start with a reference graphene type flat honey comb lattice of Si atoms, with optimized lattice parameter of 3.82 {\AA} and interatomic bond-length of 2.205 {\AA} and relax the structure to lower the energy. The simplest relaxation is a puckering that involves symmetric displacement of atoms of two triangular sublattices in opposite directions along $z$-direction. This symmetric puckering, arising from sp$^3$ mixing, has been well studied in the literature.

We go beyond the simple puckering and look at asymmetric relaxation. The structure optimizes itself by displacement of two sublattices along $z$-direction by different displacements 0.15 and 0.49 {\AA}.  Further there is a relative displacement of two sublattices along $y$-direction by 0.04 {\AA}. Hellman-Feynmann forces act on both $z$ and $y$ directions and give us the above equilibrium structure.

\begin{figure}
\includegraphics[scale=0.3]{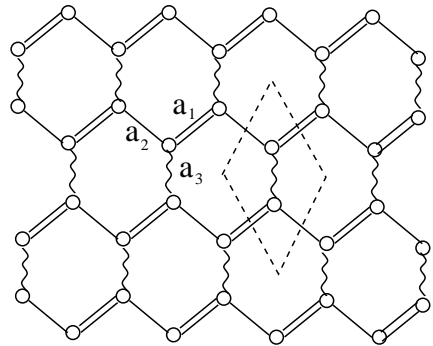}
\caption{\label{fig:kekule_str} Relaxed and asymmetrically puckered structure, from our ab initio calculation, viewed down the c-axis. Three types of bonds, double, single and wavy line have bond lengths, a$_1$ = 2.249, a$_2$ = 2.318, and a$_3$ = 2.384 {\AA} respectively.}
\end{figure}

The equilibrium structure we obtained, continues to have two atoms per unit cell. However, it has lost
the two sublattice symmetry. As illustrated in Fig.~\ref{fig:kekule_str}, we obtained three unequal nearest neighbor Si-Si bond lengths, 2.249, 2.318, and 2.384 {\AA} (experimental value of the mean is 2.24 {\AA}~\cite{aufrey06}). Top Si atoms have an optimized bond angle of 109.5$^{o}$, equal to the tetrahedral angle (sp$^3$ hybridization) , while bottom Si atoms have bond-angles of 111.9 and 114.5$^{o}$ (mixture of $sp^3$ and $sp^2$).

Our lattice parameter is 3.841 {\AA}, in close agreement with previously obtained values of 3.8 - 3.9 {\AA}~\cite{meng13}. The experimental lattice constants are in the neighborhood of 3.9 {\AA}~\cite{aufrey06}. The buckled lattice constant and bond length are slightly larger than those of planar lattice as buckling enhances $sp^{3}$ hybridization and weakens the double-bond nature between Si atoms. Further, many studies have shown that the lattice structure and buckling distance of silicene can be varied by layer-substrate interactions~\cite{vogt12,fleurence12,VB1,VB2,VB3}.

Of the three nearest neighbor bond lengths, shortest one is shorter from the mean by 3.3 $\%$. Longest one is longer from the mean by 2.4 $\%$. Figure 1 describes the valence bond order pattern arising from the above bond length changes. The ordered double bond and single bond structure resembles dimerized zig-zag polyacetylene type chains (running along x-axis with double and single bonds of lengths 2.249 and 2.318 $\AA$), that  are bonded (wavy line) along the y-axis to form a slightly elongated hexagonal lattice.

It is important to note that bond length contraction in double bonds, about the mean value, in both polyacetylene and buckyball C$_{60}$ molecules is $\approx$ 2.1 $\%$, comparable to our finding in silicene.

In addition to valence bond order, electron correlation can support 2-sublattice antiferromagnetism in a honeycomb lattice. Interestingly we find antiferromagnetism in all three cases, after energy minimization: i) planar structure, ii) symmetric puckered structure and iii) non-symmetric puckered structure.

Even though we have obtained AFM ground state with use of PBE functionals the energy difference between NM and AFM phases is very small ($\sim$ 1 meV, within accuracy of DFT calculations). This seems to point out a nearly degenerate ground states with and without magnetic order, but both having valence bond order. It is interesting that a small induced AFM moment has a substantial effect on the value of three bond lengths: i) for zero sublattice magnetization the bond lengths are (2.249, 2.318, 2.384) $\AA$ and ii) for optimized sublattice moment $\approx$ 0.015$\mu_{B}$ the bond lengths are (2.097, 2.222, 2.340) $\AA$. This large variation of bond lengths to a tiny antiferromagnetic order indicates presence of a strong spin-lattice coupling in the Mott insulating state.

\begin{figure}
\includegraphics[scale=0.4]{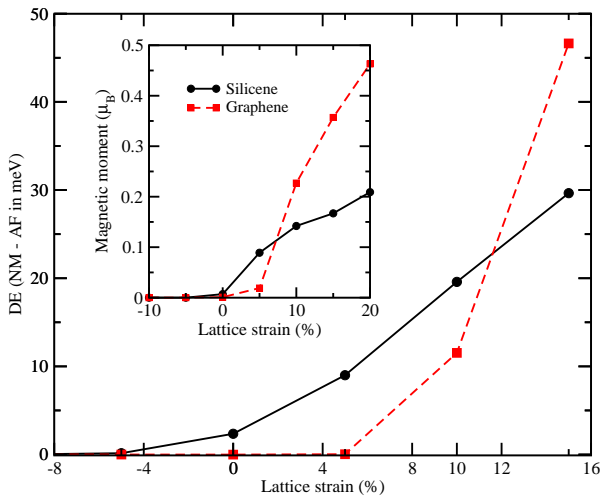}
\caption{\label{fig:AFM_gain} (Color online) Total energy difference
between non-magnetic and antiferromagnetic configurations of silicene
(black solid line) and graphene (red dashed line). Inset shows local magnetic
moment in AFM with respect to lattice strain. }
\end{figure}

Usually DFT scheme tends to overestimate the stability of delocalized states due to self-interaction errors. 
However, the experimentally observed fluorine-induced local magnetic moment in graphene was shown to be well reproduced by using HSE functional where PBE failed to do so~\cite{kim13}. Hence, we performed similar calculations using HSE functionals in NM and AFM configurations for silicene as well as graphene.

The gain in AFM energy (difference between NM and AFM total energies after optimization) with respect
to lattice strain is shown in Fig.~\ref{fig:AFM_gain}. While the AFM ordering in graphene sets in after
7\% lattice expansion, the AFM state is stable in silicene even for 5\% compressed lattice and at equilibrium. 
The gain in AFM energy steadily increases as the lattice is expanded further.
The transition to the AFM phase can be attributed to the reduction in hopping integral $t$.
We have earlier estimated that~\cite{GBsilicene} silicene has U and t values of 5 eV and 1.14 eV, respectively.
Since the $U/t$ value of silicene (4.38)  at equilibrium is greater than that of graphene
at 10\% lattice expansion (3.8), the AFM ordering occurs in silicene at equilibrium lattice itself.
This can be attributed to the larger lattice constant of Silicene than graphene, which results in lesser
overlap interaction and hence reduction in hopping integral $t$.
As the lattice is expanded, the atomic distances are uniformly increased, resulting in homogeneous increase
in $U/t$, which gives rise to steady gain in AFM energy in silicene.
The critical value of $U/t$ for the AF spin ordering in graphene has been
predicted to be 3.6 - 4.3 according to quantum Monte Carlo calculations based on Hubbard model~\cite{lee12}.
As the relative onsite Coulomb energy $U/t$ is subcritical, there is no magnetic ordering in graphene at equilibrium.

We also carried out LDA+$U$ to account for on-site Coulomb interaction. We used $U$ and $J$ values, 5 and 1 eV, respectively. When LDA+$U$ was done with PBE functional, no magnetic moment was present, similar to simple GGA calculations. Further the lattice also relaxes in a different fashion leading to bond lengths (2.310, 2.311 and 2.311) Å. The Kekule order is very weak.

States with Kekule (valence bond) order support unpaired neutral spin soliton or charge $\pm$e soliton, similar to topological excitations in RVB states and polyacetylene. They can arise in equilibrium through disorder, temperature or through substrate influence. In what follows we offer evidences for periodically ordered array of spin or charge solitons, in the background of ordered valence bonds and resonances in some of the known reconstructions of silicene grown on metallic substrates.

A variety of silicene reconstructions, $\sqrt{3} \times \sqrt{3}$, 3 $\times$ 3, $\sqrt{7} \times \sqrt{7}$ , $\sqrt{13} \times \sqrt{13}$ etc. have been discussed in the experimental literature \cite{vogt12,fleurence12,VB1,VB2,VB3}. \textit{We find that they can be interpreted as a generalized Kekule order}. A normal Kekule order or valence bond order is a lattice filled with an ordered pattern of non-overlapping dimers. We define a generalized Kekule order to contain in addition ordered i) spin or charge solitons and ii) plaquette resonances (plaquette RVB). Charge $\pm$ e (holon or doublon) could be generated through charge transfer between silicene and substrate.
\begin{figure}
\includegraphics[scale=0.3]{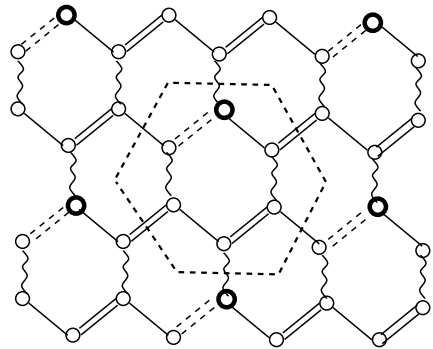}
\caption{Valence bond interpretation of $\sqrt{3} \times \sqrt{3}$ reconstruction seen in silicene on
ZrB$_2$ (1000) surface. It is a simple modification of our theoretical structure (figure 1), with a superimposed $\sqrt{3} \times \sqrt{3}$ order arising from the lattice match of a sublattice of Si atoms (dark circles)
with a sublattice of Zr atoms with 2 $\times$ 2 unit cell below.}
\end{figure}

First we consider silicene grown on ZrB$_2$ (0001) surface. It has a $\sqrt{3} \times \sqrt{3}$ reconstruction that is in coincidence with a 2 $\times$ 2 unit cell of the top layer of Zr atoms of ZrB$_2$ substrate. We interpret this structure as a modification of our structure (figure 1). The valence bond order is depicted in figure 2. There are two dypes of double bonds denoted by dotted and undotted double lines.

\begin{figure}
\includegraphics[scale=0.3]{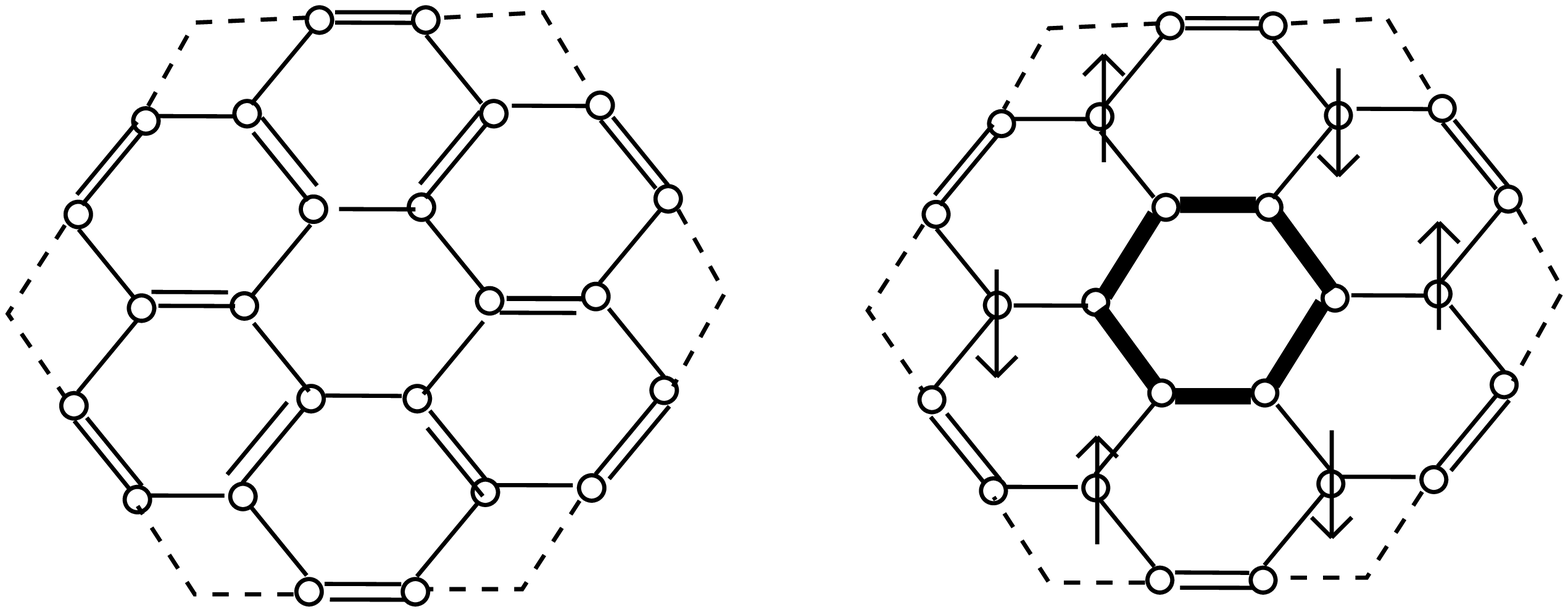}
\caption{\label{fig:Expt_inter} Two possible interpretation of the experimentally seen 3 $\times$ 3 reconstruction: a) valence bond order and b) valence bond order and a plaquette resonance (thick hexagon). The central Si hexagon in a plaquette resonance state surrounds symmetrically a Ag atom below. Sites with arrows have weakly coupled unpaired spins (spinon).}
\end{figure}

Next case is silicene grown on Ag (111) surface. In one of the reconstructed structures, a 3 $\times$ 3 unit cell of silicene coincides with a 4 $\times$ 4 unit cell of the triangular Ag surface layer. In each unit cell a planar Si hexagon surrounds a Ag atom just below. Resulting generalized Kekule pattern is shown in figure 3. Two possible Kekule patterns consistent with 3 $\times$ 3 order are:
i) ordered valence bonds (Fig.\ref{fig:Expt_inter} and ii) ordered valence bonds, weakly coupled spins (unpaired spinons) and hexagonal plaquette resonance (Fig.\ref{fig:Expt_inter}b).

When silicene lattice remains neutral, with no charge transfer from Ag substrate or ZrB$_2$ substrate, the soliton lattice forms a weakly coupled spinon (unpaired spin) sublattice, in the background of strongly coupled dimers and plaquette RVB states. The weak interaction could stabilize the spinon sublattice into a magnetically ordered state, or form a chiral or non chiral RVB State. Further, various degenerate vacua are possible for this periodic superstructure. Correspondingly various domains, topological defects and discommensurations are possible.

In the presence of charge transfer between Si lattice and substrate, the transferred charges are likely to form ordered charged solitons. This can be viewed as a holon or doublon density wave, depending on the details of the charge transfer. It will be interesting to look for this using STM and ARPES experiments.

To conclude, in this article we have, through ab-initio calculations, tried to bring out interesting manifestation of strong correlation effects in silicene. In the absence of fermi surface nesting they are potential valence bond and antiferromagnetic orders that a spin-half Mott insulator is susceptible to. It is widely believed that LDA and 
GGA are not capable of describing Mott physics, particularly spectral functions. In the present article we find that use of HSE functional in DFT is able to bring out \textit{tendencies of Mott localization}. Our result could help ways of discovering Mott physics in the DFT approach. 


As far as valence bond orders are concerned, the current experimental results of various patterns of silicene lattice reconstruction on different substrates, provide evidence for valence bond orders, spinon sub lattices and resonating valence bond subsystems. As for the long range antiferromagnetic order, it is unlikely to be stabilized, because lattice distortion will favor valence bond formation and tend to localize them.

We have indicated that increasing electron correlation effects and small sp$^3$ mixing leads to structural reconstruction. Silicene, in the absence of coulomb interactions is a semimetal with vanishing density of states. In the absence of fermi surface nesting, structural reconstructions are stabilized by Mott localization of electron and spin singlet bond formation, resulting in patterns of puckering.

Our present theoretical results, coupled with our earlier theoretical proposal \citep{GBsilicene} of a Mott insulating ground state for silicene layer, makes the system very exciting from theory and experimental point of view. A new window is likely to open into the world of novel Mott insulators, spin liquids and doped Mott insulators.

One of our predictions is presence of ordered topological defects such as spinons, holons, doublons and plaquette resonances.  It will be interesting to look for them in already existing silicene and study their spin and charge content using standard STM and spin sensitive STM.

\section{Acknowledgement}

We thank Mukul Laad and Stefano Leoni for a discussion and bringing to our attention referemces \cite{LeoniCraco,Arya}. RV is grateful to the Research Council of Norway for computing time on the Norwegian supercomputer facilities (NOTUR). RV thanks Institute of Mathematical Sciences for hospitality. GB thanks Science and Engineering Research Board (SERB, India) for the SERB Distinguished Fellowship. Additional support was provided by the Perimeter Institute for Theoretical Physics. Research at Perimeter Institute is supported by the Government of Canada through the Department of Innovation, Science and Economic Development Canada and by the Province of Ontario through the Ministry of Research, Innovation and Science.

\vskip 1cm

\begin{center}
 {\bf {\Large Supplementary Material}}
\end{center}

\vskip 0.5cm
\noindent
{\bf Valence Bond Order and Antiferromagnetism in Silicene - ab initio Results}
\vskip 0.2cm
\noindent
{R. Vidya and G. Baskaran}
\vskip 0.5cm
\begin{center}
 \textbf{Computational Details}
\end{center}

We have performed first-principles calculations using the
projected augmented plane-wave method~\cite{blochl94} implemented
in the Vienna {\it ab initio} simulation package (VASP)~\cite{vasp}. Complete
structural optimizations for free-standing silicene with a plane-wave energy cutoff of 400 eV and 550\,eV were done. The atomic geometry was optimized by force as well as
stress minimizations with convergence criteria of 10$^{-6}$\,eV
per unit cell for total energy and $\le$ 1
meV\,\rm{\,{\AA}}$^{-1}$ for forces. Exchange and correlation
effects are treated under the
generalized-gradient-approximation~\cite{perdew96} with
Perdew-Burke-Ehrenkof (PBE) functional as well as with hybrid functional. For all
studies here, the lattice parameter along $z$ direction was taken to be large enough
(15{\AA}) to ensure that no interaction remains between layers, making
them effectively isolated 2D objects.
\par

Structure optimization and electronic structure
calculation were also carried out using the Heyd-Scuseria-Ernzerhof
(HSE) hybrid functional~\cite{heyd03} to ensure proper description of electronic
structure because conventional exchange-correlation functionals are well-known to
underestimate the band-gap. We performed hybrid functional calculations for
NM and AFM spin orderings. 

In these calculations
the exact Hartree-Fock (HF) exchange energy was hybridized with the exchange-correlation
energy from the generalized gradient approximation (GGA) within the framework of a
generalized Kohn-Sham scheme to remedy the self-interaction error of GGA.
Reasonable band-gap values were empirically obtained for organic systems by taking
the mixing parameter as the inverse of the dielectric constant of the material. For graphene, the proper HF contribution could be as high as 40\%, considering the dielectric constant obtained from many-body perturbation calculations. Hence we have used a screening
parameter of 37.5\%. However, we have done a set of calculations with 25\% screening to
check its role on structural properties.

\begin{figure}[h!]
\includegraphics[scale=0.25]{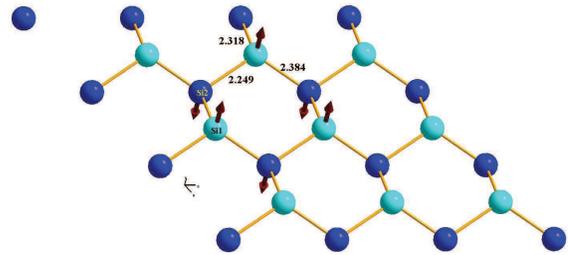}
\caption{\label{fig:magstr} (Color online) Antiferromagnetic
configuration considered in the calculation for Silicene. Interatomic
bond lengths and atom labels are given on the illustration. }
\end{figure}

The AFM structure considered for calculation is shown in Fig.~\ref{fig:magstr}. There are
two inter-penetrating triangular lattices which have opposite spin.

\end{document}